
\input phyzzx
\sequentialequations

\overfullrule=0pt
\catcode`\@=11

\def \O{{\cal O}}

\def\NP{{\it Nucl. Phys.\ }}

\def\PL{{\it Phys. Lett.\ }}

\def\Mod{{\it Mod. Phys. Lett.\ }}

\def\eqaligntwo#1{\null\,\vcenter{\openup\jot\m@th
\ialign{\strut\hfil
$\displaystyle{##}$&$\displaystyle{{}##}$&$\displaystyle{{}##}$\hfil
\crcr#1\crcr}}\,}
\catcode`\@=12
\REF\GKN{D. J. Gross, I. R. Klebanov and M. J. Newman
\journal Nucl. Phys. &B350 (1991) 621.}
\REF\ferm{D. J. Gross and I. R. Klebanov
\journal Nucl. Phys. &B352 (1991) 671.}
\REF\States{J. Goldstone, unpublished; V. G. Kac, in
{\sl Group Theoretical Methods in Physics}, Lecture Notes in Physics,
vol. 94 (Springer-Verlag, 1979).}
\REF\AMP{A. M. Polyakov, \Mod {\bf A6} (1991) 635.}
\REF\avan{J. Avan and A. Jevicki, \PL {\bf 266B} (1991) 35;
Brown preprints BROWN-HET-824 and BROWN-HET-839.}
\REF\minic{D. Minic, J. Polchinski and Z. Yang, Univ. of Texas
preprint UTTG-16-91.}
\REF\moore{G. Moore and N. Seiberg, Rutgers--Yale preprint RU-91-29,
YCTP-P19-91.}
\REF\ddmw{S. Das, A. Dhar, G. Mandal and S. Wadia,
preprints IASSNS-HEP-91/52 and 91/72.}
\REF\Ed{E. Witten, Inst. for Advanced Study preprint IASSNS-HEP-91/51.}
\REF\KP{I. R. Klebanov and A. M. Polyakov, \Mod {\bf A6} (1991) 3273.}
\REF\GKleb{D. J. Gross and I. R. Klebanov, \NP {\bf B359} (1991) 3.}
\REF\lz{B. Lian and G. Zuckerman, \PL {\bf 254B} (1991) 417;
Yale preprint YCTP-P18-91.}
\REF\bakas{I. Bakas, \PL {\bf 228B} (1989) 57.}
\REF\li{M. Li, Santa Barbara preprint UCSBTH-91-47.}
\REF\dvv{R. Dijkgraaf, E. Verlinde and H. Verlinde,
\NP {\bf B348} (1991) 435.}
\REF\Wit{E. Witten, private communication.}
\REF\klt{H. Kawai, D. Lewellen and S. H. Tye, \NP {\bf B269} (1986) 1.}
\REF\dk{P. Di Francesco and D. Kutasov, \PL {\bf 261B} (1991) 385;
Princeton preprint PUPT-1276.}
\REF\kms{D. Kutasov, E. Martinec and N. Seiberg, Rutgers
and Princeton preprint RU-91-49, PUPT-1293.}
\REF\wadia{S. Das, A. Dhar, G. Mandal and S. Wadia,
preprint IASSNS-HEP-91/79.}
\REF\ulf{U. Danielsson, Princeton preprint PUPT-1301}

\def\eqaligntwo#1{\null\,\vcenter{\openup\jot\m@th
\ialign{\strut\hfil
$\displaystyle{##}$&$\displaystyle{{}##}$&$\displaystyle{{}##}$\hfil
\crcr#1\crcr}}\,}
\catcode`\@=12

\def\half{{1\over 2}}
\Pubnum={PUPT-1302}
\date={December 1991}
\titlepage
\title{Ward Identities in Two-Dimensional String Theory.}
\author{Igor R. Klebanov }
\JHL
\abstract
I study the Ward identities of the $w_\infty$ symmetry of
the two-dimensional string theory. It is found that, not just an isolated
vertex operator, but also a number of vertex operators colliding
at a point can produce local charge non-conservation.
The structure of all such contact terms is determined. As an application,
I calculate all the non-vanishing bulk tachyon amplitudes
directly through the Ward identities for a Virasoro subalgebra
of the $w_\infty$.
\endpage

Recently, a considerable effort has been devoted to understanding
the role of discrete states in two-dimensional string theory.
In the context of the $c=1$ matrix model they were first noticed
in refs. [\GKN, \ferm], while in the continuum approach they have been known
for a long time [\States]. These physical states, that are present only
for quantized values of momentum, are the remnants of transverse
string excitations [\AMP]. They appear, somewhat surprisingly, in the
two-dimensional string model and make it quite non-trivial.
In a number of recent papers it was shown that the discrete states
generate large symmetry groups, such as the area preserving diffeomorphisms.
In the matrix model this was done in ref. [\avan-\ddmw],
while in the continuum --
in ref. [\Ed, \KP]. A better understanding of these symmetries may hold the key
to a complete solution of the theory. In this paper I derive the
symmetry Ward identities and show how they determine the
non-vanishing ``tachyon'' correlation functions.
I will restrict myself to the non-compact closed string
case and to the ``bulk'' correlation functions that satisfy the
energy sum rule [\AMP, \GKleb].

The symmetry currents were constructed
by Witten [\Ed] who first understood the important physical role of the
BRST invariant discrete states of ghost number zero [\lz]. The currents assume
the form [\Ed]
$$\eqalign{&W_{J, m}=\Psi_{J+1, m}(z)\bar\O_{J, m}(\bar z)\cr
&\bar W_{J, m}=\bar \Psi_{J+1, m}(\bar z)\O_{J, m}(z)\cr
}
\eqn\currents$$
where $J=0, \half, 1, \ldots$, and $m=-J, -J+1, \ldots, J-1, J$.
The fields $\Psi_{j, m}$ are the gravitationally dressed primary
fields of the $c=1$ matter system,
$$\Psi_{j, m}(z)=\psi_{j, m} (z)
e^{(-1+j)\phi}(z)
\eqn\eq$$
The chiral algebra of these fields was calculated in ref. [\Ed], and somewhat
more explicitly in ref. [\KP]. We will normalize fields so that
$$\Psi_{j_1, m_1}(z)\Psi_{j_2, m_2}(0)={1\over z}\ 2 (j_1 m_2-j_2 m_1)
\Psi_{j_1+j_2-1, m_1+m_2}(0) \ .
\eqn\fusion$$
The crucial point in the construction of the currents is the existence
of the BRST invariant operators $\O_{J, m}$
that form a ground ring [\Ed] generated
by
$$\eqalign{&\O_{\half, \half}=(cb+{i\over 2}\partial X-\half\partial\phi)
e^{\half(iX+\phi)}\cr
&\O_{\half, -\half}=(cb-{i\over 2}\partial X-\half\partial\phi)
e^{\half(-iX+\phi)}\cr }
$$
These operators are necessary in eq. \currents\ to balance the
right and left momenta, so that the currents are good quantum fields.
Up to BRST commutators, the ring fusion rule is [\Ed]
$$\O_{J_1, m_1}\O_{J_2, m_2}=\O_{J_1+J_2, m_1+m_2}
\eqn\ring$$
{}From the currents $W$ we can form the charges
$$Q_{J, m}={1\over 2\pi i}\oint dz W_{J, m} (z)
\ .\eqn\charges$$
Putting together eqns. \fusion, \ring\ and charges, we find the charge
algebra
$$[Q_{J_1, m_1}, Q_{J_2, m_2}] =2\{(J_1+1) m_2-(J_2+1) m_1)\}
Q_{J_1+J_2, m_1+m_2} \eqn\walgebra$$
This is known as the wedge subalgebra of $w_\infty$ [\bakas].
There are also charges $\bar Q_{J, m}$
constructed from the currents $\bar W_{J, m}$,
which satisfy the same algebra as eq. \walgebra.
For the charges to be conserved we should have
$$\bar\partial W_{J, m}=
\partial \bar W_{J, m}=0 $$
Obviously, $\partial \bar \Psi_{J+1, m}=0 $
but $\O_{J, m}$ seems to depend on $z$. However, we have
$$\partial \O=[L_{-1}, \O]=[\{Q_{BRST}, b_{-1}\}, \O]=
\{Q_{BRST}, [b_{-1}, \O]\}\ .
\eqn\contact$$
Thus, if $\O$ acts on the vacuum state, then there is no dependence on $z$.
Inside correlation functions the dependence of $\O$ on $z$ comes
only from boundary terms on the moduli space.

The symmetry Ward identities on correlation functions are encoded in
$$\VEV{\int d^2 z
\bar\partial W_{J, m} \prod_i O_i}=
\VEV{\int d^2 z \partial \bar W_{J, m} \prod_i O_i}= 0
\eqn\Ward$$
where $O_i$ are the vertex operators, three of which are
fixed and the rest integrated.
Due to the left-right symmetry on the world sheet, we only need to
consider the Ward identities due to the currents $W$. The currents
$\bar W$ give identical constraints. We will concentrate on the correlation
functions of the ``tachyon'' operators
$$T^{\pm}_k =e^{ikX+(-1\pm k)\phi}
\eqn\eq$$
where the superscript labels the chirality.

Let us begin with the role of the charges $Q_{m, m}$ and $Q_{-m, m}$.
These sets of charges
are special because they form two separate Virasoro sub-algebras
\foot{This was pointed out to me by H. Verlinde.}
$$
[Q_{n, n}, Q_{m, m}] =2(m-n) Q_{m+n, m+n}\ ,\qquad\qquad
m=0, \half, 1, \ldots
\eqn\firstval
$$
$$
[Q_{-n, n}, Q_{-m, m}] =2(m-n) Q_{-m-n, m+n}\ ,\qquad\qquad
-m=0, \half, 1, \ldots
\eqn\valgebra$$
In fact, we have only the non-negative Virasoro generators
in each case.
Note that $W_{m, m}$ carries momentum $k=m$ and Liouville
energy $\epsilon=m$. Thus, the action of $Q_{m, m}$ should shift
the momentum and energy of a tachyon by the same amount $m$, \ie\
it should convert $T^+_k$ into $T^+_{k+m}$. After an explicit calculation
we indeed find
$$ W_{m, m}(z)\ c\bar c T^+_k (0)= {1\over z} F_m(k)
\ c\bar c T^+_{k+m} (0)+\ldots
\eqn\action$$
where the first term on the right-hand side is the leading singularity.
The determination of the function $F_m(k)$ proceeds in two steps.
On the $\bar z$ side we obtain
$$\bar \O^m_{\half, \half} \bar c e^{ikX+(-1+k)\phi}
=(2k)(2k+1)\ldots (2k+2m-1)\bar c
e^{i(k+m)X+(-1+k+m)\phi}
\ ,\eqn\zebar$$
which is easily derived by a repeated application of
$$\bar \O_{\half, \half} \bar c e^{ikX+(-1+k)\phi}
=2k\bar c
e^{i(k+\half)X+(-1+k+\half)\phi}
\ .\eqn\eq$$
On the $z$ side we find
$$\Psi_{m+1, m}(z)\ c e^{ikX+(-1+k)\phi}(0)
={1\over z}{(2k+2m)!\over (2k-1)!}(-1)^{2m}
c e^{i(k+m)X+(-1+k+m)\phi}(0)
\eqn\ze$$
which can be derived using
$$\eqalign{&\psi_{m+1, m}(z)=-(2m+1)! H_-(z)e^{i(m+1)X(z)}\cr
& H_- (z)=\oint {du\over 2\pi i} e^{-iX(u+z)}\cr
 }
\eqn\eq$$
Putting together eqs. \zebar\ and \ze, we get
$$ F_m(k)=(-1)^{2m}
(2k)^2 (2k+1)^2\ldots (2k+2m-1)^2 (2k+2m)
\eqn\oldef
$$
Eq. \action\ can be simplified if we introduce specially normalized
vertex operators
$$\tilde T^+_k={\Gamma(2k)\over \Gamma(1-2k)}
T^+_k
\ .\eqn\eq$$
Then, from eqs. \action\ and \oldef\ we find simply
$$ W_{m, m}(z)\ c\bar c \tilde T^+_k (0)= {1\over z} (2k+2m)
\ c\bar c \tilde T^+_{k+m} (0)+\ldots
\eqn\newaction$$
which implies
$$[Q_{m, m}, c\bar c \tilde T^+_k]
=2(k+m)\ c\bar c \tilde T^+_{k+m}\ .\eqn\eq $$
This result is consistent with the Jacobi identity
$$[Q_{m, m}, [Q_{n, n}, c\bar c \tilde T^+_k]]
-[Q_{n, n}, [Q_{m, m}, c\bar c \tilde T^+_k]]=
[[Q_{m, m}, Q_{n, n}], c\bar c \tilde T^+_k]
\ .\eqn\eq$$

Now let us insert $Q_{m, m}$ into the correlation functions of type
$(N, 1)$, \ie\ with $N$ tachyons of chirality $+$ and 1 tachyon
of chirality $-$. Eq. \newaction\ determines how the charge
conservation is violated by each of the $+$ vertex operators.
To complete the Ward identity, we need to consider the action of
$Q_{m, m}$ on $T^-$ which, by the sum rules, carries momentum
$p=-\half (N-1)$.  This is one of the discrete momenta where
$T^-$ can be considered a special state, and we find
$$ W_{m, m}(z) \ c\bar c T^-_p (0)\sim {1\over z}
c\Psi_{-p+m, p+m} \bar X_{-p+m, p+m}
\ ,\eqn\spec$$
where $\bar X$ appears to be a mixture of $\bar c\bar \Psi$ and another state
that is not in the relative cohomology [\li]. The Ward identity
for the current $W_{m, m}$, eq. \Ward, assumes the form
$$\eqalign{&A_{N, D} (k_1, \ldots, k_N)+
(2k_1+2m)A_{N, 1}(k_1+m, k_2, \ldots, k_N)+\cr
&(2k_2+2m)A_{N, 1}(k_1, k_2+m, \ldots, k_N)+
\ldots+(2k_N+2m)A_{N, 1}(k_1, k_2, \ldots, k_N+m)=0\cr }
\eqn\oldward$$
The first term is the correlation function of the special state
$D= c\Psi\bar X$, found in eq. \spec, and $N$ tachyons of chirality
$+$; all the remaining terms are tachyon correlators of type
$(N, 1)$. Eq. \oldward\ is reminiscent of the Virasoro constraints
in $c<1$ matrix models [\dvv].

There is an interesting subtlety in the derivation
of eq. \oldward. In eq. \action\ we derived the violation of
charge conservation near a fixed vertex operator supplied with the
factor $c\bar c$. When we study a moving operator, then naively
the factor $2k+2m$ is replaced by another function, which would lead
to nonsensical results. The resolution of this problem
\foot{This puzzle was resolved by A. M. Polyakov.}
is that
$$\bar \partial W_{J, m}=
(\bar \partial \Psi_{J+1, m})\bar\O_{J, m}+
\Psi_{J+1, m}\bar\partial\bar\O_{J, m}
\ ,$$
and that the second term in the equation cannot be neglected.
In fact, by eq. \contact, its insertion reduces to boundary terms, and
each moving vertex operator
gives an extra contribution to charge non-conservation.
In this fashion, as expected, the symmetry
between the fixed and moving vertex operators is restored, and
each one introduces the factor $2k+2m$ into the Ward identity.

The Ward identities can be rephrased in a Fock space notation
for tachyons. The state created by inserting a number of tachyon operators
onto the Riemann surface will be denoted by
$|k_1, k_2, \ldots, k_N; p_1, p_2, \ldots, p_M>$,
where $k_i$ are the momenta of the $+$ tachyons, and $p_i$ are the momenta
of the $-$ tachyons.
The results above can be stated as
$$\eqalign{&Q_{m, m}|k_1, k_2, \ldots, k_N>=
(2k_1+2m)|k_1+m, k_2, \ldots, k_N>+\cr &
(2k_2+2m)|k_1, k_2+m, \ldots, k_N>+\ldots
(2k_N+2m)|k_1, k_2, \ldots, k_N+m>\ .\cr }
\eqn\eq$$
The charges $Q_{-m, m}$ act analogously on the $-$ tachyons.
The correlation functions can be written as
$$<0|{\bf S}|k_1, k_2, \ldots, k_N; p_1, p_2, \ldots, p_M>
\eqn\compact$$
where $\bf S$ is the S-matrix. A compact statement of
the Ward identities is to insert $Q_{J, m}$ into eq. \compact, and
use $[Q_{J, m}, {\bf S}]=0$ to show that the result is zero.

So far we have found that $Q_{m. m}$ acting on $+$ tachyons and
$Q_{-m, m}$ acting on $-$ tachyons
do not change the particle
number, but simply shift their momenta sequentially by a quantized amount.
However, as suggested by Witten [\Wit], in this theory one generally expects
the symmetry charges to alter the particle number. This expectation
comes true in a very interesting way. It turns out that the charge
$Q_{m+n, m}$ acts to reduce the number of $+$ tachyons by $n$,
$$\eqalign{&Q_{m+n, m}|k_1, k_2, \ldots, k_{n+1}>=
f_{m+n, m}(k_i)|k>\ ,\cr
&k=m+\sum_{i=1}^{n+1} k_i\ .\cr }
\eqn\eq$$
To show that this equation is allowed, let us count the Liouville
energy of the state on the right-hand side,
$$\epsilon=m+n+\sum_{i=1}^{n+1} (k_i-1)=-1+m+
\sum_{i=1}^{n+1} k_i=-1+k
\ .\eqn\eq$$
Thus, $k$ and $\epsilon$ of the resulting state are appropriate for
a single $+$ tachyon. Of course, it remains to show that
$f_{m+n, m}(k_i)\neq 0$, but there is no general reason for it to
vanish. We will now calculate it directly for the special case
$m=-\half$, $n=1$, using the explicit form
$$
W_{\half, -\half}=[(\partial X)^2-i\partial^2 X]
[\bar c\bar b-{i\over 2}\bar \partial X-\half\bar \partial\phi]
e^{\half(-iX+\phi)}
\ .\eqn\eq$$
We need to calculate the perturbed operator product
$$W_{\half, -\half}(z)
\ c\bar c T^+_{k_1} (0) \int d^2 w T^+_{k_2}(w, \bar w)
= {1\over z} I(k_1, k_2)
\ c\bar c T^+_{k_1+k_2-\half} (0)+\ldots
\eqn\eq$$
After all the contractions, we find an integral which
can be evaluated using the results of ref. [\klt],
$$ \eqalign{&I(k_1, k_2)=\int d^2 u |u|^{4(k_1+k_2-1)}|1-u|^{-4k_2}
\left (8k_1 k_2+2k_1(2k_1-1)(1-u)+ {2k_2(2k_2-1)\over 1-u}\right )\cr
&=2\pi (2k_1+2k_2-1)
{\Gamma(1-2k_1)\over \Gamma(2k_1)}
{\Gamma(1-2k_2)\over \Gamma(2k_2)}
{\Gamma(-1+2k_1+2k_2)\over \Gamma(2-2k_1-2k_2)}\ .\cr
} \eqn\eq$$
Changing normalization from $T^+$ to $\tilde T^+$ removes all the cumbersome
factors, and we find
$$  f_{\half, -\half}(k_1, k_2)=I(k_1, k_2)
{\Gamma(2k_1)\over \Gamma(1-2k_1)}
{\Gamma(2k_2)\over \Gamma(1-2k_2)}
{\Gamma(2-2k_1-2k_2)\over \Gamma(-1+2k_1+2k_2)}=
2\pi (2k_1+2k_2-1)
\ .$$
To summarize, we have found that
$$Q_{\half, -\half} |k_1, k_2>=
2\pi (2k_1+2k_2-1) |k_1+k_2-\half>
\ .\eqn\eq$$
We could attempt a general calculation of
$f_{m+1, m}$ along the same lines, but it is easier to use the algebra
\walgebra\ to deduce
$$ Q_{m+1, m} |k_1, k_2>={1\over 4m+3}
[Q_{\half, -\half}, Q_{m+\half, m+\half}] |k_1, k_2>=
4\pi (k_1+k_2+m) |k_1+k_2+m>
\ .\eqn\eq$$
{}From this formula we can determine the action of all the charges
$Q_{m+n, m}$. For example, for $n=2$ we may use
$$
[Q_{s+1, s}, Q_{t+1, t}] |k_1, k_2, k_3>=
4(t-s)Q_{s+t+2, s+t}|k_1, k_2, k_3> \ .\eqn\eq$$
Thus, we recursively derive
$$Q_{m+n, m}|k_1, k_2, \ldots, k_{n+1}>=
2\pi^n (n+1)!
(m+\sum_{i=1}^{n+1} k_i)
|m+\sum_{i=1}^{n+1} k_i>
\eqn\general$$
This formula is a mnemonic for how the charge conservation is violated
when all the $n+1$ vertex operators collide at a point. Thus, it is implicit
that at most one of them is of the fixed type.
If a state contains $N>n+1$ tachyons, then $Q_{m+n, m}$ acts to convert
it into a sum of $(N-n)$-particle states by turning every possible set
of $n+1$ particles into one particle according to eq. \general. If there
are $N<n+1$ tachyons, then $Q_{m+n, m}$ appears to annihilate the state.
We should keep in mind, however, that we are considering the renormalized
tachyons $\tilde T^+_k$ that, at discrete $k$, are related to
$T^+_k$ by an infinite factor.
If we do not renormalize the field, then the discrete momenta need to
be treated specially. At these momenta
there may be additional contributions to the
Ward identity, proportional to the special states. Eq.
\oldward\ is an example of this.

Finally, we note that all the
above formulae can be modified for the $-$ tachyons by a simple parity flip.
If we now introduce oscillators $a(k)$ for the renormalized $+$ tachyons,
and $b(p)$ for the renormalized $-$ tachyons, then the charges can
be represented as
$$ \eqalign{
&Q_{J, m}= 2\pi^{J-m}\int dk\prod_{i=1}^{J-m+1}\int dk_i\ ka^\dagger (k)
\prod_{s=1}^{J-m+1}a(k_s)\ \delta\left(\sum_{l=1}^{J-m+1} k_l+m-k\right)\cr
&+ (-1)^{2m}\ 2\pi^{J+m}\int dp\prod_{i=1}^{J+m+1}\int dp_i \ pb^\dagger (p)
\prod_{s=1}^{J+m+1}b(p_s)\ \delta\left(\sum_{l=1}^{J+m+1} p_l+m-p\right)
\ .\cr
}\eqn\represent
$$

Now that we have determined how the charges act on the vertex operators,
we can derive powerful constraints on the correlation functions.
As an example, I will show that the correlators of type $(N, 1)$
are completely determined by the Ward identities once we
have set $A_{2, 1}=1$ to fix the normalization of the string coupling
constant. Note that the $-$ tachyon is kept unrenormalized
because its momentum is discrete. As a warm-up, let us calculate
the 4-point function from the identity
$$<0|{\bf S}Q_{\half, -\half}|k_1, k_2, k_3; p>=0
\eqn\warmup$$
We chose $Q_{\half, -\half}$ because it can shift the momentum of
the $-$ tachyon, and it can also convert two $+$ tachyons into a single one.
This is precisely what we need to express the $(3, 1)$ amplitude in
terms of the $(2, 1)$ amplitude.

First, using the sum rules, we find that
$k_1+k_2+k_3=1$, $p=-\half$ in eq. \warmup. Then, writing out
the action of the charge, we get
$$\eqalign{&-2<0|{\bf S}|k_1, k_2, k_3; -1>+2\pi (2k_1+2k_2-1)
<0|{\bf S}|k_1+ k_2-\half, k_3; -\half>\cr &
+2\pi (2k_1+2k_3-1) <0|{\bf S}|k_1+ k_3-\half, k_2; -\half>
\cr & +2\pi (2k_2+2k_3-1) <0|{\bf S}|k_2+ k_3-\half, k_1; -\half>
=0\ .\cr }
\eqn\eq$$
It follows that
$$A_{3, 1}(k_1, k_2, k_3)=\pi (4(k_1+k_2+k_3)-3)=\pi
\ .$$
This agrees with the results of refs. [\AMP, \GKleb, \dk]
for the amplitude of the renormalized $+$ tachyons.

This procedure can be iterated. Inserting $Q_{\half, -\half}$ into
the $(4, 1)$ amplitude, we get an expression for $A_{4, 1}$ in terms of
$A_{3, 1}$. Solving it, we find $A_{4, 1}=\pi^2/2$.
Repeating the steps, it is not hard to show recursively that
$$ A_{N, 1}(k_1, k_2, \ldots, k_N)={\pi^{N-2}\over (N-2)!}
\ ,\eqn\ampl$$
in agreement with refs. [\AMP, \GKleb, \dk].
Alternatively, $A_{N, 1}$ can be found
in a single step by using
$$<0|{\bf S}Q_{-m, m}|k_1, k_2, \ldots, k_N; p>=0
\eqn\real$$
where $m=1-{N\over 2}$, $p=-\half$.
This charge can shift the momentum $p$ of the
$-$ tachyon, and it can also reduce the number
of $+$ tachyons from $N$ to 2. Writing out eq. \real, we find
$$[(N-2)!]^2 (N-1) A_{N, 1}=
\pi^{N-2} (N-1)!\biggl(N(2-N)+2(N-1)\sum_{i=1}^N k_i\biggr )=
\pi^{N-2} (N-1)!
$$
so that eq. \ampl\ again follows.
Thus, the $(N, 1)$ amplitudes are determined by the Ward identities
for the Virasoro subalgebra \valgebra. Similarly, the
$(1, N)$ amplitudes are determined
by the other Virasoro subalgebra \firstval.
\foot{Of course, the two sets of amplitudes are related by parity
inversion.} Instead of analyzing the formidable multiple
integrals, we have calculated the tachyon amplitudes in a few lines, relying
essentially only on the algebraic structure of the theory.
This encourages us to believe that all the correlation
functions are determined by the symmetries of the theory.
Further work is needed to investigate this question.

As I was finishing this work, I received a number of papers
[\kms-\ulf] addressing various closely related issues.

\ack
I am grateful to A. M. Polyakov for sharing his insight
and for many valuable suggestions. I am also indebted
to E. Witten for an inspiring comment, and to H. Verlinde
for helpful discussions.

This work was supported in part by
DOE grant DE-AC02-76WRO3072,
NSF Presidential Young Investigator Award PHY-9157482,
James S. McDonnell Foundation grant No. 91-48, and
an A. P. Sloan Foundation Research Fellowship.

\singlespace
\refout
\bye